\documentclass[twocolumn,reprint]{revtex4-1}
\usepackage[pdftex]{graphicx}
\usepackage{times}
\usepackage{bm}

\begin{document}

\title{Lowest order thermal correction to the hydrogen recombination cross section in presence of blackbody radiation}

\author{J. Triaskin}
\affiliation{Department of Physics, St.Petersburg State University, St.Petersburg, 198504, Russia
} 
\author{T. Zalialiutdinov}
\affiliation{Department of Physics, St.Petersburg State University, St.Petersburg, 198504, Russia
} 
\author{A. Anikin}
\affiliation{Department of Physics, St.Petersburg State University, St.Petersburg, 198504, Russia
} 
\author{D. Solovyev}
\email[E-mail:]{d.solovyev@spbu.ru}
\affiliation{Department of Physics, St.Petersburg State University, St.Petersburg, 198504, Russia
}

\begin{abstract}
In the present paper, the correction due to the thermal interaction of two charges to the recombination and ionization processes for the hydrogen atom is considered. The evaluation is based on a rigorous quantum electrodynamic (QED) approach within the framework of perturbation theory. The lowest-order radiative correction to the recombination/ionization cross-section is examined for a wide range of temperatures corresponding to laboratory and astrophysical conditions. The found thermal contribution is discussed both for specific states and for the total recombination and ionization coefficients.
\end{abstract}
\date{\today}

\maketitle

\section{Introduction}

The electron recombination/ionization process is widely discussed in the literature. Since the end of the 19th century, the study of this effect has found application in modern physics with the aim of a detailed description of laboratory experiments and the cosmological evolution of the early Universe. The theoretical prescription for electron recombination is precisely given within the framework of the quantum mechanical (QM) approach, which allows one to carry out the nonrelativistic evaluation (based on the solution of the Schr\"{o}dinger equation) for light atomic systems and easily extends to the relativistic case within the Dirac formalism. Recently, focusing on simple examples for the hydrogen atom, a rigorous derivation of the corresponding cross section was obtained within the framework of quantum electrodynamics (QED) \cite{Solovyev-Rec}. In particular, quantum mechanical results were obtained by considering a one-loop self-energy Feynman diagram. In addition, in \cite{Solovyev-Rec} it was demonstrated that the QED approach accommodates a thorough description of the effects induced by the blackbody radiation and, par excellence, strictly take into account the finite lifetimes of atomic levels.

One of the advantages of the QED approach is the ability to consistently take into account the radiative corrections to the recombination and ionization processes. For example, the derivation of the corresponding radiative QED corrections in the framework of the two-time Green's function method using the adiabatic S-matrix formalism can be found in \cite{shabaev-report}. Concentrating on the development of the thermal QED theory (TQED), in this paper we describe the lowest-order radiative correction that occurs when evaluating the exchange of thermal photons between two charges \cite{S-2020}. A consistent calculation of thermal corrections to the emission probabilities in hydrogen and singly ionized helium atoms were presented in \cite{ZSL-2020,ZAS-2020}, and the correction due to thermal interaction was recently evaluated in the work \cite{SZA-jpb2020}, showing its importance for the study radiation processes.

\begin{figure}[hbtp]
\caption{Feynman diagrams representing the thermal correction on the thermal interaction potential. A wavy line ($\gamma$) indicates the photon emission process, a dashed line ($\gamma_T$) corresponds to the thermal Coulomb photon exchange of a bound electron with a nucleus. The double solid line denotes the bound electron in the nucleus field (the Furry picture). Notations $i$ and $f$ represent the initial and final states of a bound electron, respectively, and $m$ corresponds to the intermediate state represented in the electron propagator.
}
\centering
\includegraphics[scale=0.2]{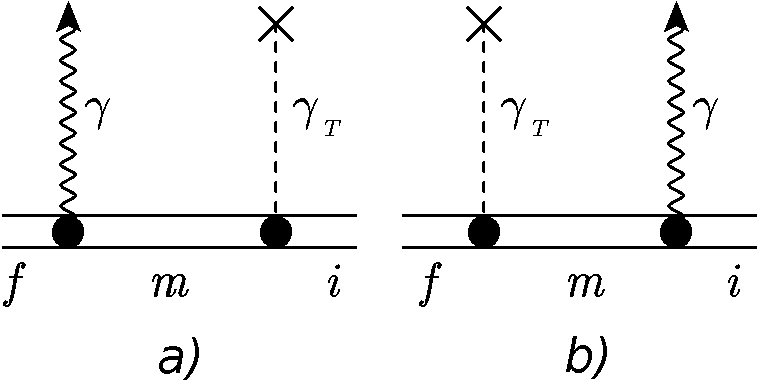}
\label{Fig-1}
\end{figure}

Adopting the formalism developed in \cite{SZA-jpb2020,Solovyev-Rec} for the vertex-type radiative thermal correction to a particular case of the radiative recombination process, we estimate the Feynman graphs shown in Fig~\ref{Fig-1}.

The process of electron transition from the initial state belonging to the continuous spectrum $i=\varepsilon$ to the bound state with the emission of a photon is considered here for the hydrogen atom placed in a heat bath. Working in nonrelativistic approximation, the wave function of the incident electron can be described as the series expansion over spherical waves \cite{Bethe, Berest, Sob, Sob2, Akhiezer}. The cross-section of recombination process, $\sigma^{\rm rec}$, can be expressed via the ionization cross-section, $\sigma^{\rm ion}$, by the detailed balance relation (in relativistic units $\hbar=c=m=1$):
\begin{eqnarray}
\label{1}
\sigma^{\rm rec}_{nl} = 2(2l+1)\sigma^{\rm ion}_{nl}\frac{k^2}{p^2},
\end{eqnarray}
where $k$ is the momentum of the emitted photon, $p\equiv|\vec{p}|$ is the incident electron momentum and $nl$ is the principal quantum number and orbital momentum of the bound atomic state, respectively. The corresponding QED derivation of the cross section for the radiative recombination process using the one-loop self-energy correction can be found in \cite{Solovyev-Rec}.

\section{Thermal vertex correction to the recombination process}
\label{SE-V}
To obtain the lowest-order thermal correction to the recombination cross section, it is convenient to use the adiabatic $S$-matrix formalism for reducible Feynman graphs (Fig.~\ref{Fig-1}), when each interaction vertex contains an additional exponential factor $\mathrm{exp}(-\eta|t|)$. The exponential pre-factor, however, is not necessarily needed at the top of the thermal interaction indicated by the cross in these diagrams. The S-matrix element corresponding to Fig.~\ref{Fig-1}a) is
\begin{eqnarray}
\label{2}
S^{(3)}_{\eta} = (-\mathrm{i} e)^2 i Ze \int d^4x_1 d^4x_2 d^4x_3 \bar{\psi}_f(x_1)\gamma^\mu A_\mu(x_1) \qquad
\\
\nonumber
\times
e^{-\eta |t_1|}S(x_1,x_2)
e^{-\eta |t_2|} \gamma^\nu 
D^\beta_{\nu\lambda}(x_2,x_3)
 j^\lambda(x_3)\psi_i(x_2),\qquad
\end{eqnarray}
where integration is extended over space-time variables $x_i$ which denote the spatial position vector $\vec{r}$ and the time variable $t$. The Dirac matrices are denoted as $\gamma^{\mu}$, where $\mu$ takes the values $\mu=(0,1,2,3)$, $\psi_a(x)=\psi_a({\vec{r}})e^{-\mathrm{i}E_a t}$ is the one-electron Dirac wave function, $\bar{\psi_a}$ is the Dirac conjugated wave function and $j^\sigma(x)$ is the four-dimensional nuclear current.

The standard electron propagator defined as the vacuum-expectation value of the time-ordered product of electron-positron field operators can be represented in terms of an eigenmode decomposition with respect to one-electron eigenstates \citep{Akhiezer,LabKlim}:
\begin{eqnarray}
\label{3}
S(x_1,x_2)
= \frac{\mathrm{i}}{2\pi}\int\limits_{-\infty}^{\infty}d\omega e^{-\mathrm{i}\omega(t_1-t_2)}\sum\limits_n\frac{\psi_n({\vec{r}_1})\bar{\psi}_n({\vec{r}_2})}{\omega-E_n(1-\mathrm{i}0)},\;\;\;
\end{eqnarray}
where summation runs over the entire Dirac spectrum. 
The photon wave function, $A_{\mu}(x)$, is
\begin{eqnarray}
\label{4}
A_{\mu}(x)=\sqrt{\frac{2\pi}{\omega}}e^{(\lambda)}_{\mu}e^{ik_{\mu}x^{\mu}}.
\end{eqnarray}
Here $e^{(\lambda)}_{\mu}$ are the components of the photon polarization 4-vector, $x_{\mu}$ is the space-time 4-vector, $k_{\mu}$ is the photon momentum 4-vector with the space vector $\vec{k}$ and photon frequency $\omega=|\vec{k}|$. Using the transversality condition $\gamma_{\mu}e^{(\lambda)}_{\mu}=\vec{e}\vec{\alpha}$ ($\vec{e}$ is a transverse  space vector of the photon polarization), the wave function for the emitted/absorbed real photon  takes the form:
\begin{eqnarray}
\label{5}
\vec{A}(x) = \sqrt{\frac{2\pi}{\omega}}\,\vec{e}e^{i(\vec{k}\vec{r}-\omega t)}
 \equiv \sqrt{\frac{2\pi}{\omega}}\,e^{-i\omega t}\,\vec{A}(\vec{k},\vec{r}).
\end{eqnarray}

\begin{figure}[hbtp]
	\centering
	\includegraphics[scale=0.15]{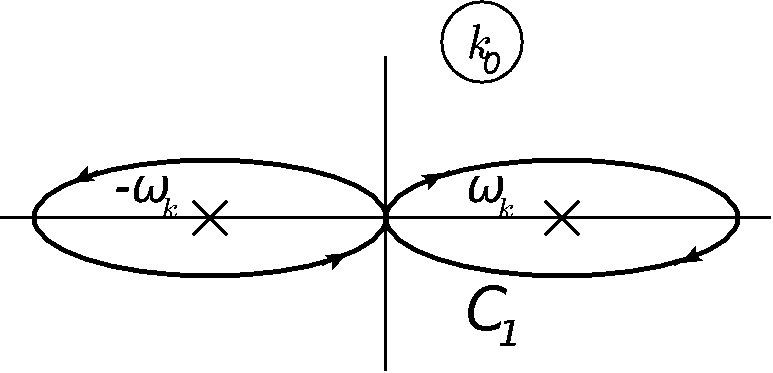} 
	\caption{Integration contour $C_1$ in $k_0$ plane. Arrows on the contour define the pole-bypass rule. The poles $\pm\omega_k$ are denoted with $\times$ marks.}
	\label{Fig-2}
\end{figure}

The thermal part of photon propagator was found in \cite{S-2020} in the form:
\begin{eqnarray}
\label{6}
D_{\lambda\sigma}^\beta(x_2 x_3) = 4\pi \int\limits_{C_1}\frac{d^4k}{(2\pi)^4}\frac{e^{ik(x_2-x_3)}}{k^2}n_\beta(|\vec{k}|),
\end{eqnarray}
where $k^2\equiv k_0^2-\vec{k}^2$, $n_\beta(|\vec{k}|)$ represents the Planck distribution function $(\mathrm{exp}(\beta|\vec{k}|)-1)^{-1}$, $\beta = 1/(k_B T)$, $k_B$ is the Boltzmann constand and $T$ is temperature in Kelvin. The notation $C_1$ in Eq. (\ref{5}) denotes the integration in $k_0$ plane over the contour shown in Fig.~\ref{Fig-2}.

At first one can integrate over the $d^4x_3$ variables in Eq. (\ref{2}), which leads to the four-dimensional Fourier transform of the nuclear current $j^\sigma(k)$. For the point-like nucleaus within the static limit it can be simplified to $j^\sigma(k) = j^0(k) = 2\pi \delta(k_0)\rho(\vec{k}) = 2\pi \delta(k_0)$. Then the arising $\delta$-function leads to the doubled three-dimensional Fourier transform of the function $n_\beta(|\vec{k}|)/\vec{k}^2$. A rigorous derivation of the remaining integrals can be found in \cite{S-2020}, which gives rise to the thermal Coulomb potential. 

Then, the $S$-matrix element, Eq. (\ref{2}), can be found as
\begin{eqnarray}
\label{7}
S^{(3)}_{\eta} = -4\pi Ze^3 \int d^4x_1 d^4x_2\bar{\psi}_f(x_1)\gamma^\mu A_\mu(x_1)e^{-\eta |t_1|}  
\\
\nonumber
\times
S(x_1,x_2) e^{-\eta |t_2|} \int\frac{d^3k}{(2\pi)^3}\frac{e^{i\vec{k}\vec{r}_2}}{\vec{k}^2}n_\beta(|\vec{k}|)\psi_i(x_2)
.
\end{eqnarray}
It should be noted here that the same expression could immediately be written in the thermal Coulomb gauge and must be regularized at $|\vec{k}|\equiv \kappa\rightarrow 0$, see \cite{S-2020,SZA-jpb2020}.
The subsequent evaluation of the Feynman graphs in Fig.~\ref{Fig-1} we omit for brevity (the corresponding calculations completely repeat the content of \cite{SZA-jpb2020}).

According to \cite{SZA-jpb2020}, the regularized thermal correction to the emission probability is reduced to
\begin{eqnarray}
\label{8}
\Delta W_{if}^{{\rm rad}} = \frac{4 Ze^4\zeta(3)}{9\pi^2\beta^3}\langle i|\vec{\alpha}\vec{A}|f\rangle \omega_{if} d\vec{\nu}\times  \qquad
\\
\nonumber
\left[ 
\mathop{{\sum}'}\limits_m \frac{\langle f | \vec{\alpha}\vec{A}^* | m \rangle\langle m | r^2 | i \rangle}{E_i-E_m}
 + 
 \mathop{{\sum}'}\limits_m \frac{\langle f | r^2 | m \rangle\langle m | \vec{\alpha}\vec{A}^* | i \rangle}{E_f-E_m} 
\right.
\\
\left.
\nonumber
+ \frac{1}{2}\frac{\langle f | \vec{\alpha}\vec{A}^* | i \rangle\langle i | r^2 | i \rangle }{\omega_{if}  } - \frac{1}{2}\frac{\langle f |r^2 | f \rangle\langle f | \vec{\alpha}\vec{A}^* | i \rangle }{\omega_{if}  } \right],
\end{eqnarray}
where $\zeta(3)$ is the Riemann zeta function. 
The recombination cross section $d\sigma$ is related to the transition probability by the relation $d\sigma = dW/j$, where $j=\upsilon$ is the particle flux density per unit volume ($\upsilon$ is the velocity of particles equal to the speed of light for photons).

One of the conclusions following from the result Eq. (\ref{8}) is that matrix elements containing scalar operator $r^2$ preserves the parity of the state, i.e. the matrix element $(r^2)_{nm}$ is nonzero for states with the same orbital angular momentum due to the orthogonality property. Thus, further integration over the angles of the momentum $\vec{p}$ represented in the electron wave function for the continuum state can be performed in an ordinary manner using the orthogonality property for the Legendre polynomials, $P_l(\cos\theta)$:
\begin{eqnarray}
\label{9}
\int d\theta_{\vec{p}} P_{l'}(\cos\theta_{\vec{p}\vec{r'}})P_{l}(\cos\theta_{\vec{p}\vec{r}}) = \frac{4\pi}{2l+1}P_l(\cos\theta_{\vec{r}\vec{r'}}),\qquad
\end{eqnarray}
and recurrent formula
\begin{eqnarray}
\label{10}
x P_l(x) = \frac{(l+1)}{(2l+1)}P_{l+1}(x)+\frac{l}{(2l+1)}P_{l-1}(x).
\end{eqnarray}
The wave function for the state from the continuum with the energy $\varepsilon=p^2/2$ can be written in the form:
\begin{eqnarray}
\label{11}
\psi_p = \frac{1}{2p}\sum\limits_{l=0}^\infty \mathrm{i}^l(2l+1) e^{\mathrm{i}\delta_l}R_{pl}(r)P_l(\cos\theta_{\vec{p}\vec{r}}),
\end{eqnarray}
where $R_{pl}(r)$ is the radial part of the wave function, and the phase factor $\delta_l$ can be omitted as immaterial for our purposes.

The result for the electric dipole photon emission is well known and leads to
\begin{eqnarray}
\label{12}
\int d\theta_{\vec{p}} d\theta_{\vec{r}} d\theta_{\vec{r'}} \langle \varepsilon l'|\vec{r}|nl\rangle \langle nl|\vec{r'}| m l' \rangle = \qquad
\\
\nonumber
l\frac{(4\pi)^2}{2l+1} I_{pl-1;nl}I_{nl;ml-1} + (l+1)\frac{(4\pi)^2}{2l+1} I_{pl+1;nl}I_{nl;ml+1},
\end{eqnarray}
which holds for $n=m$, and $l'=l\mp 1$, respectively.
Here 
\begin{eqnarray}
\label{13}
I_{pl';nl} = 
\int\limits_0^\infty dr\, r^3 R_{nl}(r)R_{pl'}(r),
\end{eqnarray}
Analytical representation of the radial wave functions of discrete $R_{nl}(r)$ and the continuum $R_{pl'}(r)$ states for the hydrogen atom can be found in textbooks \cite{Berest, Akhiezer}. Then radial integrals of the type Eq. (\ref{13}) are usually calculated employing the Gordon formula, see, for example, \cite{Karzas,Boardman,Burgess}. The expression (\ref{12}) is written for the first term in Eq. (\ref{8}) and easily adapts to the second one.


Combining all the results, the final expression for recombination to an arbitrary bound $nl$ state can be written as
\begin{eqnarray}
\label{14}
\Delta\sigma_{nl} = 
\frac{64 Ze^4\zeta(3)}{9(2l+1)\beta^3} 
l_>\left[-\frac{1}{2}\; I_{pl';nl}\,R_{nl;nl}I_{nl;pl'}+
\right.
\nonumber
\\
\left.
+\sum\limits_{m\atop (m\neq \varepsilon)}\frac{E_n - E_m}{E_\varepsilon-E_m}  I_{pl';nl}\,I_{nl;ml'}R_{ml';pl'}  + \qquad
\right.
\\
\nonumber
\left.
+\sum\limits_{m\atop (m\neq n)}\frac{E_m-E_\varepsilon}{E_n-E_m}\; I_{pl';nl}\,I_{ml;pl'}R_{nl;ml}
\right](E_\varepsilon - E_n)^2,
\end{eqnarray}
where $l_>={\rm max}(l,l')$ and the expression (\ref{14}) consists of two contributions with $l'=l-1$ and $l'=l+1$ according to (\ref{12}). Pointing out that the last but one term in Eq. (\ref{8}) is a correction to the wave function of the continuum state, it can be excluded from consideration, see \cite{shabaev-report}. Here we have introduced the notation:
\begin{eqnarray}
\label{15}
R_{nl;pl'} = 
\int\limits_0^\infty dr\, r^4 R_{nl}(r) R_{pl'}(r) = \qquad
\\
\nonumber
\frac{2^{l+l'+1}p^{l'}n^{-l-2}}{[(2l+1)!]^2}\sqrt{\frac{(n+l)!}{(n-l-1)!}}\left[\frac{8\pi p}{1-e^{-\frac{2\pi}{p}}}\right]^{1/2}\qquad
\\
\nonumber
\times 
\prod\limits_{s=1}^{l'}\sqrt{s^2+\frac{1}{p^2}}\int\limits_0^\infty dr\, r^{4+l+l'}\, e^{-\frac{r}{n}-i p r}\times\qquad
\\
\nonumber
 F\left(-n+l+1,2l+2,\frac{2r}{n}\right)F\left(\frac{i}{p}+l'+1,2l'+2,2i p r\right)
\end{eqnarray}

The integral (\ref{15}) (as well as (\ref{13}), that leads to Gordon's formula) can be calculated analytically using the derivative with respect to the parameter before $r$ in the exponent, the multiplicity of the derivative is determined by reducing it to a tabular integral:
\begin{eqnarray}
\label{16}
\int\limits_0^\infty dt\, t^{c-1}e^{-s t}\,_1F_1\left(a;c;t\right)\,_1F_1\left(\alpha;c;\lambda t\right) = \qquad
\\
\nonumber
\frac{(c-1)!}{(s-1)^a(s-\lambda)^\alpha}s^{a+\alpha-c}\,_2F_1\left(a,\alpha;c;\frac{\lambda}{(s-1)(s-\lambda)}\right).
\end{eqnarray}
Here $_1F_1$ is the confluent hypergeometric functions of the first kind and $_2F_1$ is the Gauss's hypergeometric functions. As well as the first contribution in Eq. (\ref{14}) is given with $R_{nl;nl} = \frac{n^2}{2}(5n^2+1-3l(l+1))$.

The analytical result for $R_{nl;nl}$ shows an impetuous growth with an increase of $n$, which makes us conclude the significance of the correction Eq. (\ref{14}) for highly excited states. Nonetheless, as pointed out in \cite{SZA-jpb2020} the approximation $r\ll 1$ is valid for low-lying states and may be violated for Rydberg states. The legitimacy of using such an approximation is dictated by the series expansion of the potential found in \cite{S-2020} in the vicinity $\frac{r}{\beta}\ll 1$. In \cite{SZA-jpb2020}, it was found (see Table IV there) that the calculations of the full form for the thermal potential and approximated by the $r^2$ contribution deviate starting from $n=20$ at 300 K and $n=10$ at 3000 K. However, we now found that the $r/\beta$ thermal potential argument was parametrized incorrectly (the $\alpha$ was omitted). Numerical values corresponding to the correction of the lowest order \cite{SZA-jpb2020} were recalculated with the correct scaling and are listed in Table~\ref{tab:comparison}.
\begin{center}
\begin{table}[hbtp]
\caption{Numerical values of energy shifts $\Delta E^{\beta}_{A}=\langle A |V^\beta(r)| A\rangle$ for different atomic states $A$ at temperatures $T=300$ K (upper line) and $T=3000$ K (lower line) in hydrogen atom. The first column shows the considered state $(n_{A},l_{A})$. In the second column the energy shift is calculated with approximate potential $V^\beta(r)$ given by Eqs. (38) and (52) in \cite{SZA-jpb2020}. In the third column energy shift is calculated with potential $V^\beta(r)$ given by Eq. (51) in \cite{SZA-jpb2020}. All values are in Hz. }
\begin{tabular}{l c c}
\hline
\hline
$ (n_{A},l_{A}) $  & $\Delta E^{\beta}_{n_{A}l_{A}} $, Eq. (38)  & $ \Delta E^{\beta}_{n_{A}l_{A}}  $, Eq. (51) \\
\hline
(1,0)  & $-3.36$ & $-3.36$\\
       & $-3.36\times 10^{3}$&  $-3.36\times 10^{3}$\\
(2,0)  & $-46.98$ & $-46.98$ \\
       & $-4.698\times 10^4$ & $-4.698\times 10^4$\\
(10,0) & $-2.80\times 10^4 $ & $-2.80\times 10^4$\\
       & $-2.80\times 10^7$& $-2.80\times 10^7$\\      
(10,9) & $-1.29\times 10^4$ & $-1.29\times 10^4$\\
       & $-1.29\times 10^7$& $-1.29\times 10^7$\\   
(20,0) & $-4.48\times 10^5$ & $-4.48\times 10^5$\\
       & $-4.48\times 10^8 $& $-4.47\times 10^8$ \\ 
(20,19)& $-1.93\times 10^5$ & $-1.93 \times 10^5$\\
       & $-1.93\times 10^8$& $ -1.93\times 10^8$\\  
(100,0)& $-2.80\times 10^8$ & $ -2.78\times 10^8$\\
       & $-2.80\times 10^{11}$& $-2.78\times 10^{11}$\\   
(100,99)& $-1.14\times 10^8$& $-1.13 \times 10^8$\\
       & $ -1.14\times 10^{11} $& $ -9.171\times 10^{10} $\\      
(200,0)& $-4.47\times 10^9$ & $-4\times 10^9$\\
       & $ -4.47\times 10^{12} $& $  -3.72\times 10^{11} $\\     
(200,99)& $-1.80\times 10^9$& $-1.73 \times 10^{9}$\\
       & $ -1.80\times 10^{12} $& $  -5.06\times 10^{11} $\\                         
\hline
\hline
\end{tabular}
\label{tab:comparison}
\end{table}
\end{center}
As a result, it turns out that there is no deviation up to $n\approx 100$ at such temperatures. The recalculated Table~V from \cite{SZA-jpb2020} is given below:
\begin{widetext}
\begin{center}
\begin{table}[hbtp]
\caption{Transition rates and thermal corrections at $ T=300 $ K to one-photon electric dipole transitions between highly excited states due to the thermal energy shift, see Eqs. (53), (54) and Table~V in \cite{SZA-jpb2020}. All values are given in s$^{-1}$.}
\begin{tabular}{l l c c c c}
\hline
\hline
$n_{i},l_{i}$ & $n_{f},l_{f}$ & $ W_{if} $ & $ \Delta W^{\rm ind}_{if} $ & $\Delta W^{{\rm v}}_{if}$ & $\Delta W^{{\rm v, ind}}_{if}$ \\
\hline
$(10,9)$ & $(9,8)$ & $ 1.320\times 10^{4} $ & $ 5.419\times 10^{3} $ & $ 2.213\times 10^{-5} $ &  $ 2.811\times 10^{-6} $\\

$ (50,1) $ & $(49,0)  $ & $ 2.682 $& $ 3.077\times 10^{2} $ & $ 1.998\times 10^{-4} $ & $ 1.524\times 10^{-2} $\\

$ (50,49) $ & $(49,48)  $ & $ 7.137\times 10^{-1} $& $81.861$ & $ 2.190\times 10^{-5} $ & $ 1.671\times 10^{-3} $\\

 $ (70,1) $ & $(69,0)  $ & $ 4.840\times 10^{-1} $& $1.541\times10^{2}$ & $2.759\times 10^{-4} $ & $ 5.852\times 10^{-2}$\\

$ (70,69) $ & $(69,68)  $ & $ 9.369\times 10^{-2} $& $29.830$ & $ 2.186\times 10^{-5} $ & $ 4.636\times 10^{-3}$\\

  $ (100,1) $ & $(99,0)  $ & $ 7.953\times 10^{-2} $& $ 74.387$ & $ 3.858\times 10^{-4} $ & $ 2.407\times 10^{-1} $\\

$ (100,99) $ & $(99,98)  $ & $ 1.093\times 10^{-2} $& $ 10.221$ & $ 2.175\times 10^{-5} $ & $ 1.356\times 10^{-2} $\\

\hline

\hline
\hline
\end{tabular}
\label{tab:2d}
\end{table}
\end{center}
\end{widetext}

\section{Recombination and ionization coefficients}
The thermal correction to the effective cross-sections evaluated in the previous section allows one to define the corresponding correction to the recombination and ionization coefficients \cite{Sob2}. The rate of recombination to the $n$-th level due to the spontaneous recombination processes, $\alpha_{nl}$, is given by
\begin{eqnarray}
\label{26}
\alpha_{nl}=\int\limits_{0}^{\infty}\sigma_{nl}^{\rm rec}f(v)vdv,
\end{eqnarray}
where $\sigma_{nl}^{\rm rec}$ represents the spontaneous recombination cross section, $f(v)$ is the Maxwell-Boltzmann distribution function with the velocity of incident electrons $v$ ($v=p$ in our units):
\begin{eqnarray}
\label{27}
f(v) dv = 4\pi \left(\frac{1}{2\pi k_B T}\right)^{3/2} v^2 e^{-\frac{v^2}{2k_B T_{e}}} dv\,.
\end{eqnarray}
The presence of the Maxwell-Boltzmann distribution function in the recombination coefficient restricts the magnitude of the incident electron momentum $p$. The typical speed can be estimated as $p^2\sim 2k_B T\ll 1$ up to $T\sim 10^5$ K what justifies the used non-relativistic approximation.

The similar expression can be written for the stimulated recombination coefficient
\begin{eqnarray}
\label{28}
\alpha_{nl}^\beta=\int\limits_{0}^{\infty}\sigma_{nl}^{\rm rec,\beta}f(v)vdv,
\end{eqnarray}
 and the total recombination coefficient is
\begin{eqnarray}
\label{29}
\alpha^{\rm total}\equiv\alpha_A=\sum\limits_{nl}\alpha_{nl},
\end{eqnarray}
where index $A$ corresponds to the so-called case A when the coefficient $\alpha^{\rm total}$ includes the direct recombination process to the ground state, while case B in astrophysical researches excludes this process. 

Recently the influence of finite lifetimes on the stimulated transition rates in hydrogen and helium atoms has been studied in \cite{SLP-QED,Zal-2018hel,Zal-19}, while this effect for bound-free transitions is described in detail in \cite{Solovyev-Rec}. In the latter case, the numerical calculations become much more complicated when summing over $nl$ for the recombination/ionization coefficients due to the presence of the Lorentz factor. The effect of finite lifetimes itself in the recombination process reaches a level of few percent of the 'ordinary' stimulated transitions, leveling out at high temperatures and large values of $nl$. Although the corresponding widths of atomic levels can be taken into account here, we will leave it and focus on numerical calculations of the corresponding well-known spontaneous and stimulated rates. The latter can be expressed, see \cite{Sob,Sob2}, as
\begin{eqnarray}
\label{30}
\sigma_{nl}^{\rm rec,\beta} = \sigma_{nl}^{\rm rec}n_\beta(\varepsilon + E_{nl}),
\end{eqnarray}
where $E_{nl}$ is the ionization potential of the $nl$ state. 


The corrections to the partial spontaneous and stimulated recombination coefficients ($\Delta\alpha_{nl}$ and $\Delta\alpha^\beta_{nl}$, respectively), partial ionization coefficient ($\Delta\beta_{nl}$), that we are interested in can also be calculated using Eqs. (\ref{1}), (\ref{26}), (\ref{28})-(\ref{30}). The corresponding numerical results  for the $1s$ and $2s$ states in the hydrogen atom are given in Table~\ref{tab:1} separately for each three summand in Eq. (\ref{14}). It should be noted here that the calculations are well converged upon summation over the intermediate spectrum $m$, which were carried out by direct summation of each individual state $ml<\varepsilon$ to $m=100$. The values listed in Table~\ref{tab:1} are guaranteed to be within five digits.
\begin{widetext}
\begin{center}
\begin{table}
\caption{Thermal corrections to the partial recombination and ionization coefficients for spontaneous and stimulated processes for the $1s$ and $2s$ states at different temperatures. The coefficients $\alpha_{nl}$ are calculated using Eq. (\ref{14}), the first, second and third contributions are denoted as $\Delta\alpha_{1s}^{(1)}$, $\Delta\alpha_{1s}^{(2)}$, $\Delta\alpha_{1s}^{(3)}$, respectively. Values with index $ \beta $ denote corresponding stimulated recombination corrections. Summation over $m$ in Eq. (\ref{14}) was performed in the range $m\in[1,100]$, which guarantees the given numbers in the table. The correction to the partial ionization coefficient $\Delta\beta_{nl}$ is given as a total contribution and coincides with the sum of $\Delta\alpha_{nl}^{(1)}$, $\Delta\alpha_{nl}^{(2)}$, $\Delta\alpha_{nl}^{(3)}$, $\Delta\alpha_{nl}^{\beta,(1)}$, $\Delta\alpha_{and}^{\beta,(2)}$ and $\Delta\alpha_{and}^{\beta,(3)}$, as it should be according to the detailed balance. All values are given in $\mathrm{m}^3\mathrm{s}^{-1}$.}	
\label{tab:1}
\begin{tabular}{l c c c c c c}
\hline
\hline
{} & $T=300$ K &  $T=1000$ K &  $T=3000$ K &  $T=5000$ K &  $T=10000$ K &  $T=20000$ K\\
\hline
\hline
$\alpha_{1s}$ & $9.4939\times 10^{-19}$ & $ 5.1848\times 10^{-19} $ & $ 2.9688\times 10^{-19} $ & $2.2812\times 10^{-19}$ & $ 1.5819\times 10^{-19} $  & $ 1.0787\times 10^{-19} $\\

$\alpha_{1s}^\beta$ & $0.0$ & $6.9968\times 10^{-88}$ & $2.0781\times 10^{-42}$ & $2.2263\times 10^{-33}$ & $1.1211\times 10^{-26}$ & $2.0858\times 10^{-23}$\\

$\Delta\alpha_{1s}^{(1)}$ & $-3.3362\times 10^{-29}$ & $-6.6434\times 10^{-28}$ & $-1.0148\times 10^{-26}$ & $-3.5689\times 10^{-26}$ & $-1.9282\times 10^{-25}$ & $-1.0049\times 10^{-24}$\\

$\Delta\alpha_{1s}^{\beta,(1)}$ & $0.0$ & $-8.9930\times 10^{-97}$ & $-7.1673\times 10^{-50}$ & $-3.5334\times 10^{-40}$ & $-1.4032\times 10^{-32}$ & $-2.0339\times 10^{-28}$\\

$\Delta\alpha_{1s}^{(2)}$ & $-1.4502\times 10^{-24}$ & $-1.0971\times 10^{-23}$ & $-6.3683\times 10^{-23}$ & $-1.4172\times 10^{-22}$ & $-4.1445\times 10^{-22}$ & $-1.1997\times 10^{-21}$\\

$\Delta\alpha_{1s}^{\beta,(2)}$ & $0.0$ & $-2.6489\times 10^{-92}$ & $-8.3235\times 10^{-46}$ & $-2.6099\times 10^{-36}$ & $-5.5717\times 10^{-29}$ & $-4.3721\times 10^{-25}$\\

$\Delta\alpha_{1s}^{(3)}$ & $-3.3162\times 10^{-29}$ & $-6.4671\times 10^{-28}$ & $-9.4444\times 10^{-27}$ & $-3.2380\times 10^{-26}$ & $-1.6872\times 10^{-25}$ & $-8.5139\times 10^{-25}$\\

$\Delta\alpha_{1s}^{\beta,(3)}$ & $0.0$ & $-8.9365\times 10^{-97}$ & $-6.8742\times 10^{-50}$ & $-3.3163\times 10^{-40}$ & $-1.2731\times 10^{-32}$ & $-1.7797\times 10^{-28}$\\

\hline
$\Delta\beta_{1s}$ & $-1.4503\times 10^{-24}$ & $-1.0972\times 10^{-23}$ & $-6.3702\times 10^{-23}$ & $-1.4178\times 10^{-22}$ & $-4.1479\times 10^{-22}$ & $-1.2019\times 10^{-21}$\\
\hline

$\alpha_{2s}$ & $1.3919\times 10^{-19}$ & $7.6117\times 10^{-20}$ & $4.3716\times 10^{-20}$ & $3.3664\times 10^{-20}$ & $2.3419\times 10^{-20}$ & $1.5998\times 10^{-20}$\\

$\alpha_{2s}^\beta$ & $5.02703\times 10^{-77}$ & $2.7449\times 10^{-37}$ & $4.2385\times 10^{-26}$ & $6.3229\times 10^{-24}$ & $2.3283\times 10^{-22}$ & $1.2711\times 10^{-21}$\\

$\Delta\alpha_{2s}^{(1)}$ & $-1.9237\times 10^{-29}$ & $-3.8311\times 10^{-28}$ & $-5.6857\times 10^{-27}$ & $-1.9496\times 10^{-26}$ & $-1.0001\times 10^{-25}$ & $-4.8333\times 10^{-25}$\\

$\Delta\alpha_{2s}^{\beta,(1)}$ & $-6.9737\times 10^{-87}$ & $-1.3982\times 10^{-45}$ & $-5.6939\times 10^{-33}$ & $-3.8470\times 10^{-30}$ & $-1.0792\times 10^{-27}$ & $-4.3657\times 10^{-26}$\\

$\Delta\alpha_{2s}^{(2)}$ & $-1.8429\times 10^{-26}$ & $-1.3955\times 10^{-24}$ & $-8.1121\times 10^{-24}$ & $-1.8065\times 10^{-23}$ & $-5.2879\times 10^{-23}$ & $-1.5317\times 10^{-22}$ \\

$\Delta\alpha_{2s}^{\beta,(2)}$ & $-1.0928\times 10^{-82}$ & $-9.0071\times 10^{-42}$ & $-1.4704\times 10^{-29}$ & $-6.4156\times 10^{-27}$ & $-1.0055\times 10^{-24}$ & $-2.4067\times 10^{-23}$ \\

$\Delta\alpha_{2s}^{(3)}$ & $-2.6134\times 10^{-28}$ & $-5.0962\times 10^{-27}$ & $-7.4361\times 10^{-26}$ & $-2.5487\times 10^{-25}$ & $-1.3296\times 10^{-24}$ & $-6.7418\times 10^{-24}$ \\

$\Delta\alpha_{2s}^{\beta,(3)}$ & $-3.5101\times 10^{-52}$ & $-1.8839\times 10^{-44}$ & $-7.5179\times 10^{-32}$ & $-5.0406\times 10^{-29}$ & $-1.4108\times 10^{-26}$ & $-5.7965\times 10^{-25}$ \\

 \hline
$\Delta\beta_{2s}$ & $-1.8458\times 10^{-25}$ & $-1.4010\times 10^{-24}$ & $-8.1921\times 10^{-24}$ & $-1.8346\times 10^{-23}$ & $-5.5330\times 10^{-23}$ & $-1.8508\times 10^{-22}$ \\
  \hline
  \hline
\end{tabular}
\end{table}
\end{center}
\end{widetext}

The numerical results in Table~\ref{tab:1} show mostly insignificant contributions to the partial coefficients $\alpha_{1s(2s)}$, $\alpha^\beta_{1s(2s)}$ and $\beta_{1s(2s)}$. However, according to the discussion in the end of the previous section and the definition Eq. (\ref{29}), summation over $nl$ leads to an increase in the heat correction for the total coefficients $\alpha_A$, $\alpha^\beta_A$ and $\beta_A$ to such an extent that the summation result does not converge. Situations in which the same pattern occurs were discussed in \cite{Hummer-1988,Boschan}. A stocktaking of the effects limiting the divergent partition sum $\sum\limits_{nl} (2l+1) n_{nl}^{({\rm Boltzmann})}$ is described in detail in \cite{Hummer-1988}. The simplified model in our case is as follows. The probability $w_n$ that the state $n$ is not destroyed by the mixing thermal interaction corresponding to the matrix element $(r^2)_{ab}$ between two arbitrary states $a$ and $b$ should be inserted into the sum over $nl$ states in Eq. (\ref{29}). Then, according to Eq. (\ref{8}), we compare the thermal correction $\Delta E^\beta_{nl}\sim\beta^{-3}n^2(5n^2+1-3l(l+1))$, see \cite{S-2020}, with Lamb shift scaled $\Delta E_L\sim 1.24214\times 10^{-6}n^{-3}$ for the $ns(l=0)$ state in atomic units \cite{LabKlim}. We solve equation $\Delta E_L=\Delta E^\beta_{ns}$ for a specified temperature, which gives the same results if the partition function $\mathrm{exp}(-(\Delta E^\beta_{ns})/\Delta E_L)$ equaled to $e^{-1}$. The result can be written as
\begin{eqnarray}
\label{model}
n^* &=& \frac{1.14026}{(k_B T)^\frac{3}{7}},
\\
\nonumber
w_n &=& e^{-(\frac{n}{n^*})^7} \approx e^{-0.399 (k_B T)^3 n^7}
\end{eqnarray}
in atomic units.

Still one should take into account the thermal energy shift for the energy levels of the atom in the unperturbed cross section. This can be done by modifying the unperturbed cross section by replacing $E_a \rightarrow E_a + \Delta E_a^\beta$. Then, it can be found that the third and fourth contributions in Eq. (\ref{8}) (or the first one in Eq. (\ref{14})) are canceled out by this replacement, and contributions proportional to the cube and the square of $\Delta E_a^\beta$ remain. However, these corrections are of the next order in $\alpha$, so we omit their further calculations. 


Below are the results of numerical calculations of the total ionization and recombination coefficients and thermal corrections to them. The case B can be easily obtained by the subtraction of corresponding values of $\alpha_{1s}$ from $\alpha_A$, see Table~\ref{tab:1}. Numerical values of the total coefficients $\alpha_A$, $\alpha_A^\beta$, $\Delta\alpha_A$, $\beta_A$ and $\Delta\beta_A$ are collected in Table~\ref{tab:2} for different temperatures. The values are obtained by direct summation of partial coefficients with the partition function Eq. (\ref{model}) up to $n,m=100$.
\begin{widetext}
\begin{center}
\begin{table}
\caption{The corrections to the total recombination and ionization coefficients for spontaneous and stimulated processes for the case A at different temperatures. All values are given in $\mathrm{m}^3\mathrm{s}^{-1}$.}	\label{tab:2}
\begin{tabular}{ c | c  c  c  c  c  c  c}
\hline
{} & $T=300$ K &  $T=700$ K & $T=1000$ K &  $T=3000$ K &  $T=5000$ K &  $T=10000$ K &  $T=20000$ K\\
\hline
$\alpha_A$ & $4.32385\times 10^{-18}$ & $2.52126\times 10^{-18} $ & $2.00071\times 10^{-18}$ & $9.63800\times 10^{-19}$ & $6.78908\times 10^{-19}$  & $4.16397\times 10^{-19}$ & $2.50652\times 10^{-19}$\\

$\alpha^{\beta}_A$ & $2.15163\times 10^{-18}$ & $1.72895\times 10^{-18}$ & $1.56064\times 10^{-18}$ & $1.10529\times 10^{-18}$ & $9.29960\times 10^{-19}$ & $7.28372\times 10^{-19}$ & $5.65045\times 10^{-19}$\\
\hline



$\Delta\alpha_A$ & $-2.29004\times 10^{-20}$ & $-1.41894\times 10^{-20}$ & $-1.16107\times 10^{-20}$ & $-6.26605\times 10^{-21}$ & $-5.04184\times 10^{-21}$ & $-2.74662\times 10^{-21}$ & $-2.64351\times 10^{-21}$\\

$\Delta\alpha_A^\beta$ & $2.56355\times 10^{-21}$ & $1.50305\times 10^{-21}$ & $1.16062\times 10^{-21}$ & $4.99707\times 10^{-22}$ & $8.75126\times 10^{-23}$ & $2.46134\times 10^{-22}$ & $-6.53582\times 10^{-23}$\\

\hline
\hline





$\beta_A$  & $6.47549\times 10^{-18}$ & $4.25021\times 10^{-18}$ & $3.56135\times 10^{-18}$ & $2.06909\times 10^{-18}$ & $1.60887\times 10^{-18}$ & $1.14477\times 10^{-18}$ & $8.15697\times 10^{-19}$\\

$\Delta\beta_A$  & $-2.03369\times 10^{-20}$ & $-1.26864\times 10^{-20}$ & $-1.04501\times 10^{-20}$ & $-5.76635\times 10^{-21}$ & $-4.95433\times 10^{-21}$ & $-2.50048\times 10^{-21}$ & $-2.70887\times 10^{-21}$\\



 \hline
  \hline
\end{tabular}
\end{table}
\end{center}
\end{widetext}

\section{Discussion and conclusions}

The numerical results obtained in this work for the thermal correction Eq. (\ref{14}) for specific $1s$ and $2s$ states are given in Table~\ref{tab:1}. One can find an increasing value of the correction with elevating temperature. In particular, considering the recombination process at room temperature $300$ K, the thermal contribution is $-1.4503 \times 10^{-24}$, whereas the spontaneous recombination coefficient for the $1s$ state is about $10^{-18}$. This relation is valid for the $2s$ state, which leads to the conclusion that for a ratio of about $10^{-6}$ this correction is rather insignificant in laboratory experiments. The opposite case corresponds to higher temperatures. For example, at a temperature $20\,000$ K, the thermal correction to the recombination cross-section reaches a level of $1.1\%$ with respect to the spontaneous one and is two orders of magnitude larger than the stimulated recombination coefficient $\alpha_{1s}^\beta$. The relative value of the order of $1.1\%$ with respect to spontaneous recombination into the $2s$ state is retained, but the thermal correction is an order of magnitude less than the stimulated coefficient. Thus, one can expect a significant contribution of the thermal correction to the total (summed over all $nl$ states) recombination coefficient. Moreover, following directly from the discussion presented above, see also \cite{S-2020}, the increasing value of the correction with the principal quantum number $n$ sets the need for such a calculation.

Performing a direct summation over $nl$ of the thermal correction Eq. (\ref{14}) results in a diverging contribution. To 'streamline' this, we followed the procedure described in \cite{Hummer-1988,Boschan}, where physical conditions are discussed in detail. According to \cite{Hummer-1988} the probability $w_n$ that the state $n$ is not destroyed by the mixing thermal interaction should be introduced, limiting the divergent partition sum. The numerical results of the summation with the probability $w_n$, Eq. (\ref{model}), are listed in Table~\ref{tab:2}. 

In particular, as follows from Table~\ref{tab:2}, the thermal correction to the total recombination coefficient is about $0.3\%$ at any temperature. As a matter, this value can be compared with the achieved accuracy of astrophysical experiments aimed at studying the recombination of the early Universe. 
Then, considering the thermal effect giving by Eq. (\ref{14}) in the astrophysical context of the recombination of the early universe, the fitting formula for the total recombination coefficient (the same as in \cite{Seager_2000}):
\begin{eqnarray}
\label{31}
\alpha^S_B=10^{-19}\frac{a\, t^b}{1+c\, t^d}{\rm m^3 s^{-1}},
\end{eqnarray}
can be found with the parameters $a = 4.4648$, $b = -0.6092$, $c = 0.7470$ and $d = 0.5049$ ($t = T_M/10^4$ K) instead of $a=4.309$, $b= -0.6166$, $c=0.6703$, and $d=0.5300$ known from \cite{Pequignot,Verner}. We used the data from Table~\ref{tab:2} to find the estimate in the modification of the ionization fraction. As in \cite{Solovyev-Rec}, a change in the coefficients $a$, $b$, $c$ and $d$ can lead to $0.2\%$ contribution to the ionization fraction of the primordial plasma, repeating the effect of finite lifetimes of atomic states (the contribution decreases with increasing temperature and more significant for low temperatures). However, such a seemingly insignificant contribution is of interest  for further planned experimental data and is highlighted by the constantly produced new data with unprecedented precision \cite{Glover}.

\section*{Acknowledgments}
This work was supported by Russian Science Foundation (Grant No. 17-12-01035).

\bibliography{mybibfile}

\begin{thebibliography}{24}%
\makeatletter
\providecommand \@ifxundefined [1]{%
 \@ifx{#1\undefined}
}%
\providecommand \@ifnum [1]{%
 \ifnum #1\expandafter \@firstoftwo
 \else \expandafter \@secondoftwo
 \fi
}%
\providecommand \@ifx [1]{%
 \ifx #1\expandafter \@firstoftwo
 \else \expandafter \@secondoftwo
 \fi
}%
\providecommand \natexlab [1]{#1}%
\providecommand \enquote  [1]{``#1''}%
\providecommand \bibnamefont  [1]{#1}%
\providecommand \bibfnamefont [1]{#1}%
\providecommand \citenamefont [1]{#1}%
\providecommand \href@noop [0]{\@secondoftwo}%
\providecommand \href [0]{\begingroup \@sanitize@url \@href}%
\providecommand \@href[1]{\@@startlink{#1}\@@href}%
\providecommand \@@href[1]{\endgroup#1\@@endlink}%
\providecommand \@sanitize@url [0]{\catcode `\\12\catcode `\$12\catcode
  `\&12\catcode `\#12\catcode `\^12\catcode `\_12\catcode `\%12\relax}%
\providecommand \@@startlink[1]{}%
\providecommand \@@endlink[0]{}%
\providecommand \url  [0]{\begingroup\@sanitize@url \@url }%
\providecommand \@url [1]{\endgroup\@href {#1}{\urlprefix }}%
\providecommand \urlprefix  [0]{URL }%
\providecommand \Eprint [0]{\href }%
\providecommand \doibase [0]{http://dx.doi.org/}%
\providecommand \selectlanguage [0]{\@gobble}%
\providecommand \bibinfo  [0]{\@secondoftwo}%
\providecommand \bibfield  [0]{\@secondoftwo}%
\providecommand \translation [1]{[#1]}%
\providecommand \BibitemOpen [0]{}%
\providecommand \bibitemStop [0]{}%
\providecommand \bibitemNoStop [0]{.\EOS\space}%
\providecommand \EOS [0]{\spacefactor3000\relax}%
\providecommand \BibitemShut  [1]{\csname bibitem#1\endcsname}%
\let\auto@bib@innerbib\@empty
\bibitem [{\citenamefont {Solovyev}\ \emph {et~al.}(2019)\citenamefont
  {Solovyev}, \citenamefont {Zalialiutdinov}, \citenamefont {Anikin},
  \citenamefont {Triaskin},\ and\ \citenamefont {Labzowsky}}]{Solovyev-Rec}%
  \BibitemOpen
  \bibfield  {author} {\bibinfo {author} {\bibfnamefont {D.}~\bibnamefont
  {Solovyev}}, \bibinfo {author} {\bibfnamefont {T.}~\bibnamefont
  {Zalialiutdinov}}, \bibinfo {author} {\bibfnamefont {A.}~\bibnamefont
  {Anikin}}, \bibinfo {author} {\bibfnamefont {J.}~\bibnamefont {Triaskin}}, \
  and\ \bibinfo {author} {\bibfnamefont {L.}~\bibnamefont {Labzowsky}},\ }\href
  {\doibase 10.1103/PhysRevA.100.012506} {\bibfield  {journal} {\bibinfo
  {journal} {Phys. Rev. A}\ }\textbf {\bibinfo {volume} {100}},\ \bibinfo
  {pages} {012506} (\bibinfo {year} {2019})}\BibitemShut {NoStop}%
\bibitem [{\citenamefont {Shabaev}(2002)}]{shabaev-report}%
  \BibitemOpen
  \bibfield  {author} {\bibinfo {author} {\bibfnamefont {V.}~\bibnamefont
  {Shabaev}},\ }\href {\doibase https://doi.org/10.1016/S0370-1573(01)00024-2}
  {\bibfield  {journal} {\bibinfo  {journal} {Physics Reports}\ }\textbf
  {\bibinfo {volume} {356}},\ \bibinfo {pages} {119 } (\bibinfo {year}
  {2002})}\BibitemShut {NoStop}%
\bibitem [{\citenamefont {Solovyev}(2020)}]{S-2020}%
  \BibitemOpen
  \bibfield  {author} {\bibinfo {author} {\bibfnamefont {D.}~\bibnamefont
  {Solovyev}},\ }\href {\doibase https://doi.org/10.1016/j.aop.2020.168128}
  {\bibfield  {journal} {\bibinfo  {journal} {Annals of Physics}\ }\textbf
  {\bibinfo {volume} {415}},\ \bibinfo {pages} {168128} (\bibinfo {year}
  {2020})}\BibitemShut {NoStop}%
\bibitem [{\citenamefont {Zalialiutdinov}\ \emph
  {et~al.}(2020{\natexlab{a}})\citenamefont {Zalialiutdinov}, \citenamefont
  {Solovyev},\ and\ \citenamefont {Labzowsky}}]{ZSL-2020}%
  \BibitemOpen
  \bibfield  {author} {\bibinfo {author} {\bibfnamefont {T.}~\bibnamefont
  {Zalialiutdinov}}, \bibinfo {author} {\bibfnamefont {D.}~\bibnamefont
  {Solovyev}}, \ and\ \bibinfo {author} {\bibfnamefont {L.}~\bibnamefont
  {Labzowsky}},\ }\href {\doibase 10.1103/PhysRevA.101.052503} {\bibfield
  {journal} {\bibinfo  {journal} {Phys. Rev. A}\ }\textbf {\bibinfo {volume}
  {101}},\ \bibinfo {pages} {052503} (\bibinfo {year}
  {2020}{\natexlab{a}})}\BibitemShut {NoStop}%
\bibitem [{\citenamefont {Zalialiutdinov}\ \emph
  {et~al.}(2020{\natexlab{b}})\citenamefont {Zalialiutdinov}, \citenamefont
  {Anikin},\ and\ \citenamefont {Solovyev}}]{ZAS-2020}%
  \BibitemOpen
  \bibfield  {author} {\bibinfo {author} {\bibfnamefont {T.}~\bibnamefont
  {Zalialiutdinov}}, \bibinfo {author} {\bibfnamefont {A.}~\bibnamefont
  {Anikin}}, \ and\ \bibinfo {author} {\bibfnamefont {D.}~\bibnamefont
  {Solovyev}},\ }\href {\doibase 10.1103/PhysRevA.102.032204} {\bibfield
  {journal} {\bibinfo  {journal} {Phys. Rev. A}\ }\textbf {\bibinfo {volume}
  {102}},\ \bibinfo {pages} {032204} (\bibinfo {year}
  {2020}{\natexlab{b}})}\BibitemShut {NoStop}%
\bibitem [{\citenamefont {Solovyev}\ \emph {et~al.}(2021)\citenamefont
  {Solovyev}, \citenamefont {Zalialiutdinov},\ and\ \citenamefont
  {Anikin}}]{SZA-jpb2020}%
  \BibitemOpen
  \bibfield  {author} {\bibinfo {author} {\bibfnamefont {D.}~\bibnamefont
  {Solovyev}}, \bibinfo {author} {\bibfnamefont {T.}~\bibnamefont
  {Zalialiutdinov}}, \ and\ \bibinfo {author} {\bibfnamefont {A.}~\bibnamefont
  {Anikin}},\ }\href {\doibase 10.1088/1361-6455/abd2d0} {\bibfield  {journal}
  {\bibinfo  {journal} {Journal of Physics B: Atomic, Molecular and Optical
  Physics}\ }\textbf {\bibinfo {volume} {54}},\ \bibinfo {pages} {095001}
  (\bibinfo {year} {2021})}\BibitemShut {NoStop}%
\bibitem [{\citenamefont {Bethe}\ and\ \citenamefont {Salpeter}(1957)}]{Bethe}%
  \BibitemOpen
  \bibfield  {author} {\bibinfo {author} {\bibfnamefont {H.~A.}\ \bibnamefont
  {Bethe}}\ and\ \bibinfo {author} {\bibfnamefont {E.}~\bibnamefont
  {Salpeter}},\ }\href {\doibase 10.1007/978-1-4613-4104-8} {\emph {\bibinfo
  {title} {Quantum Mechanics of One- and Two-Electron Atoms}}}\ (\bibinfo
  {publisher} {Springer Berlin Heidelberg},\ \bibinfo {year}
  {1957})\BibitemShut {NoStop}%
\bibitem [{\citenamefont {Berestetskii}\ \emph {et~al.}(1982)\citenamefont
  {Berestetskii}, \citenamefont {Lifshits},\ and\ \citenamefont
  {Pitaevskii}}]{Berest}%
  \BibitemOpen
  \bibfield  {author} {\bibinfo {author} {\bibfnamefont {V.}~\bibnamefont
  {Berestetskii}}, \bibinfo {author} {\bibfnamefont {E.}~\bibnamefont
  {Lifshits}}, \ and\ \bibinfo {author} {\bibfnamefont {L.}~\bibnamefont
  {Pitaevskii}},\ }\href@noop {} {\emph {\bibinfo {title} {Quantum
  Electrodynamics}}}\ (\bibinfo  {publisher} {Oxford: Butterworth-Heinemann},\
  \bibinfo {year} {1982})\BibitemShut {NoStop}%
\bibitem [{\citenamefont {Sobel'man}(1972)}]{Sob}%
  \BibitemOpen
  \bibfield  {author} {\bibinfo {author} {\bibfnamefont {I.~I.}\ \bibnamefont
  {Sobel'man}},\ }\href@noop {} {\emph {\bibinfo {title} {Introduction to the
  Theory of Atomic Spectra}}}\ (\bibinfo  {publisher} {Pergamon},\ \bibinfo
  {year} {1972})\BibitemShut {NoStop}%
\bibitem [{\citenamefont {Sobelman}(1996)}]{Sob2}%
  \BibitemOpen
  \bibfield  {author} {\bibinfo {author} {\bibfnamefont {I.~I.}\ \bibnamefont
  {Sobelman}},\ }\href@noop {} {\emph {\bibinfo {title} {Atomic Spectra and
  Radiative Transitions}}},\ Vol.\ \bibinfo {volume} {2nd Edition}\ (\bibinfo
  {publisher} {Springer-Verlag Berlin Heidelberg New York},\ \bibinfo {year}
  {1996})\BibitemShut {NoStop}%
\bibitem [{\citenamefont {Akhiezer}\ and\ \citenamefont
  {Berestetskii}(1965)}]{Akhiezer}%
  \BibitemOpen
  \bibfield  {author} {\bibinfo {author} {\bibfnamefont {A.~I.}\ \bibnamefont
  {Akhiezer}}\ and\ \bibinfo {author} {\bibfnamefont {V.~B.}\ \bibnamefont
  {Berestetskii}},\ }\href@noop {} {\emph {\bibinfo {title} {Quantum
  Electrodynamics}}}\ (\bibinfo  {publisher} {Wiley-Interscience, New York},\
  \bibinfo {year} {1965})\BibitemShut {NoStop}%
\bibitem [{\citenamefont {Labzowsky}\ \emph {et~al.}(1993)\citenamefont
  {Labzowsky}, \citenamefont {Klimchitskaya},\ and\ \citenamefont
  {Dmitriev}}]{LabKlim}%
  \BibitemOpen
  \bibfield  {author} {\bibinfo {author} {\bibfnamefont {L.}~\bibnamefont
  {Labzowsky}}, \bibinfo {author} {\bibfnamefont {G.}~\bibnamefont
  {Klimchitskaya}}, \ and\ \bibinfo {author} {\bibfnamefont {Y.}~\bibnamefont
  {Dmitriev}},\ }\href@noop {} {\emph {\bibinfo {title} {Relativistic Effects
  in the Spectra of Atomic Systems}}}\ (\bibinfo  {publisher} {Institute of
  Physics Publishing},\ \bibinfo {year} {1993})\BibitemShut {NoStop}%
\bibitem [{\citenamefont {{Karzas}}\ and\ \citenamefont
  {{Latter}}(1961)}]{Karzas}%
  \BibitemOpen
  \bibfield  {author} {\bibinfo {author} {\bibfnamefont {W.~J.}\ \bibnamefont
  {{Karzas}}}\ and\ \bibinfo {author} {\bibfnamefont {R.}~\bibnamefont
  {{Latter}}},\ }\href {\doibase 10.1086/190063} {\bibfield  {journal}
  {\bibinfo  {journal} {Astrophysical Journal, Supplement Series}\ }\textbf
  {\bibinfo {volume} {6}},\ \bibinfo {pages} {167} (\bibinfo {year}
  {1961})}\BibitemShut {NoStop}%
\bibitem [{\citenamefont {Boardman}(1964)}]{Boardman}%
  \BibitemOpen
  \bibfield  {author} {\bibinfo {author} {\bibfnamefont {W.~J.}\ \bibnamefont
  {Boardman}},\ }\href {\doibase 10.1086/190101} {\bibfield  {journal}
  {\bibinfo  {journal} {Astrophysical Journal Supplement}\ }\textbf {\bibinfo
  {volume} {9}},\ \bibinfo {pages} {185} (\bibinfo {year} {1964})}\BibitemShut
  {NoStop}%
\bibitem [{\citenamefont {{Burgess}}(1965)}]{Burgess}%
  \BibitemOpen
  \bibfield  {author} {\bibinfo {author} {\bibfnamefont {A.}~\bibnamefont
  {{Burgess}}},\ }\href@noop {} {\bibfield  {journal} {\bibinfo  {journal}
  {Memoirs of the Royal Astronomical Society}\ }\textbf {\bibinfo {volume}
  {69}},\ \bibinfo {pages} {1} (\bibinfo {year} {1965})}\BibitemShut {NoStop}%
\bibitem [{\citenamefont {Solovyev}\ \emph {et~al.}(2015)\citenamefont
  {Solovyev}, \citenamefont {Labzowsky},\ and\ \citenamefont
  {Plunien}}]{SLP-QED}%
  \BibitemOpen
  \bibfield  {author} {\bibinfo {author} {\bibfnamefont {D.}~\bibnamefont
  {Solovyev}}, \bibinfo {author} {\bibfnamefont {L.}~\bibnamefont {Labzowsky}},
  \ and\ \bibinfo {author} {\bibfnamefont {G.}~\bibnamefont {Plunien}},\ }\href
  {\doibase 10.1103/PhysRevA.92.022508} {\bibfield  {journal} {\bibinfo
  {journal} {Phys. Rev. A}\ }\textbf {\bibinfo {volume} {92}},\ \bibinfo
  {pages} {022508} (\bibinfo {year} {2015})}\BibitemShut {NoStop}%
\bibitem [{\citenamefont {{Zalialiutdinov}}\ \emph {et~al.}(2018)\citenamefont
  {{Zalialiutdinov}}, \citenamefont {{Solovyev}},\ and\ \citenamefont
  {{Labzowsky}}}]{Zal-2018hel}%
  \BibitemOpen
  \bibfield  {author} {\bibinfo {author} {\bibfnamefont {T.}~\bibnamefont
  {{Zalialiutdinov}}}, \bibinfo {author} {\bibfnamefont {D.}~\bibnamefont
  {{Solovyev}}}, \ and\ \bibinfo {author} {\bibfnamefont {L.}~\bibnamefont
  {{Labzowsky}}},\ }\href {\doibase 10.1088/1361-6455/aa8e6e} {\bibfield
  {journal} {\bibinfo  {journal} {Journal of Physics B Atomic Molecular
  Physics}\ }\textbf {\bibinfo {volume} {51}},\ \bibinfo {eid} {015003}
  (\bibinfo {year} {2018})},\ \Eprint {http://arxiv.org/abs/1709.07210}
  {arXiv:1709.07210 [physics.atom-ph]} \BibitemShut {NoStop}%
\bibitem [{\citenamefont {Zalialiutdinov}\ \emph {et~al.}(2019)\citenamefont
  {Zalialiutdinov}, \citenamefont {Solovyev}, \citenamefont {Labzowsky},\ and\
  \citenamefont {Plunien}}]{Zal-19}%
  \BibitemOpen
  \bibfield  {author} {\bibinfo {author} {\bibfnamefont {T.}~\bibnamefont
  {Zalialiutdinov}}, \bibinfo {author} {\bibfnamefont {D.}~\bibnamefont
  {Solovyev}}, \bibinfo {author} {\bibfnamefont {L.}~\bibnamefont {Labzowsky}},
  \ and\ \bibinfo {author} {\bibfnamefont {G.}~\bibnamefont {Plunien}},\ }\href
  {\doibase 10.1103/PhysRevA.99.012502} {\bibfield  {journal} {\bibinfo
  {journal} {Phys. Rev. A}\ }\textbf {\bibinfo {volume} {99}},\ \bibinfo
  {pages} {012502} (\bibinfo {year} {2019})}\BibitemShut {NoStop}%
\bibitem [{\citenamefont {{Hummer}}\ and\ \citenamefont
  {{Mihalas}}(1988)}]{Hummer-1988}%
  \BibitemOpen
  \bibfield  {author} {\bibinfo {author} {\bibfnamefont {D.~G.}\ \bibnamefont
  {{Hummer}}}\ and\ \bibinfo {author} {\bibfnamefont {D.}~\bibnamefont
  {{Mihalas}}},\ }\href {\doibase 10.1086/166600} {\bibfield  {journal}
  {\bibinfo  {journal} {\apj}\ }\textbf {\bibinfo {volume} {331}},\ \bibinfo
  {pages} {794} (\bibinfo {year} {1988})}\BibitemShut {NoStop}%
\bibitem [{\citenamefont {Boschan}\ and\ \citenamefont
  {Biltzinger}(1996)}]{Boschan}%
  \BibitemOpen
  \bibfield  {author} {\bibinfo {author} {\bibfnamefont {P.}~\bibnamefont
  {Boschan}}\ and\ \bibinfo {author} {\bibfnamefont {P.}~\bibnamefont
  {Biltzinger}},\ }\href@noop {} {\bibfield  {journal} {\bibinfo  {journal}
  {Astronomy and Astrophysics}\ }\textbf {\bibinfo {volume} {336}},\ \bibinfo
  {pages} {1} (\bibinfo {year} {1996})}\BibitemShut {NoStop}%
\bibitem [{\citenamefont {Seager}\ \emph {et~al.}(2000)\citenamefont {Seager},
  \citenamefont {Sasselov},\ and\ \citenamefont {Scott}}]{Seager_2000}%
  \BibitemOpen
  \bibfield  {author} {\bibinfo {author} {\bibfnamefont {S.}~\bibnamefont
  {Seager}}, \bibinfo {author} {\bibfnamefont {D.~D.}\ \bibnamefont
  {Sasselov}}, \ and\ \bibinfo {author} {\bibfnamefont {D.}~\bibnamefont
  {Scott}},\ }\href {\doibase 10.1086/313388} {\bibfield  {journal} {\bibinfo
  {journal} {The Astrophysical Journal Supplement Series}\ }\textbf {\bibinfo
  {volume} {128}},\ \bibinfo {pages} {407} (\bibinfo {year}
  {2000})}\BibitemShut {NoStop}%
\bibitem [{\citenamefont {{Pequignot}}\ \emph {et~al.}(1991)\citenamefont
  {{Pequignot}}, \citenamefont {{Petitjean}},\ and\ \citenamefont
  {{Boisson}}}]{Pequignot}%
  \BibitemOpen
  \bibfield  {author} {\bibinfo {author} {\bibfnamefont {D.}~\bibnamefont
  {{Pequignot}}}, \bibinfo {author} {\bibfnamefont {P.}~\bibnamefont
  {{Petitjean}}}, \ and\ \bibinfo {author} {\bibfnamefont {C.}~\bibnamefont
  {{Boisson}}},\ }\href@noop {} {\bibfield  {journal} {\bibinfo  {journal}
  {Astron. Astrophys.}\ }\textbf {\bibinfo {volume} {251}},\ \bibinfo {pages}
  {680} (\bibinfo {year} {1991})}\BibitemShut {NoStop}%
\bibitem [{\citenamefont {{Verner}}\ and\ \citenamefont
  {{Ferland}}(1996)}]{Verner}%
  \BibitemOpen
  \bibfield  {author} {\bibinfo {author} {\bibfnamefont {D.~A.}\ \bibnamefont
  {{Verner}}}\ and\ \bibinfo {author} {\bibfnamefont {G.~J.}\ \bibnamefont
  {{Ferland}}},\ }\href {\doibase 10.1086/192284} {\bibfield  {journal}
  {\bibinfo  {journal} {Astr. J. Supp.}\ }\textbf {\bibinfo {volume} {103}},\
  \bibinfo {pages} {467} (\bibinfo {year} {1996})},\ \Eprint
  {http://arxiv.org/abs/astro-ph/9509083} {arXiv:astro-ph/9509083 [astro-ph]}
  \BibitemShut {NoStop}%
\bibitem [{\citenamefont {Glover}\ \emph {et~al.}(2014)\citenamefont {Glover},
  \citenamefont {Chluba}, \citenamefont {Furlanetto}, \citenamefont
  {Pritchard},\ and\ \citenamefont {Savin}}]{Glover}%
  \BibitemOpen
  \bibfield  {author} {\bibinfo {author} {\bibfnamefont {S.~C.}\ \bibnamefont
  {Glover}}, \bibinfo {author} {\bibfnamefont {J.}~\bibnamefont {Chluba}},
  \bibinfo {author} {\bibfnamefont {S.~R.}\ \bibnamefont {Furlanetto}},
  \bibinfo {author} {\bibfnamefont {J.~R.}\ \bibnamefont {Pritchard}}, \ and\
  \bibinfo {author} {\bibfnamefont {D.~W.}\ \bibnamefont {Savin}}\ }(\bibinfo
  {publisher} {Academic Press},\ \bibinfo {year} {2014})\ pp.\ \bibinfo {pages}
  {135--270}\BibitemShut {NoStop}%
\end{thebibliography}%

\end{document}